\documentclass{pnastwo}
\pdfoutput=1

\usepackage{graphics}

\usepackage{amssymb,amsfonts,amsmath}

\newcommand{\lcdm}{$\Lambda$CDM}
\newcommand{\kms}{\,{\rm km}\,{\rm s}^{-1}}
\newcommand{\cgm}{\,{\rm cm^2}\,{\rm g}^{-1}}
\newcommand{\apj}{Astrophys. J.}
\newcommand{\aj}{Astron. J.}
\newcommand{\apjl}{Astrophys. J. Lett.}
\newcommand{\nat}{Nature}
\newcommand{\prd}{Phys Rev D}
\newcommand{\mnras}{Mon. Not. Roy. Astr. Soc.}
\newcommand{\jcap}{JCAP}

\contributor{Submitted to Proceedings
of the National Academy of Sciences of the United States of America}
\url{www.pnas.org/cgi/doi/10.1073/pnas.0709640104}
\copyrightyear{2008}
\issuedate{Issue Date}
\volume{Volume}
\issuenumber{Issue Number}

\begin{document}



\title{Cold dark matter: controversies on small scales}





\author{David H. Weinberg\affil{1}{Ohio State University, Columbus, OH, USA},
James S. Bullock\affil{2}{University of California at Irvine, Irvine, CA, USA},
Fabio Governato\affil{3}{University of Washington, Seattle, WA, USA},
Rachel Kuzio de Naray\affil{4}{Georgia State University, Atlanta, GA, USA}, and
Annika H. G. Peter\affil{1}{}\affil{2}{}
}

\contributor{Submitted to Proceedings of the National Academy of Sciences
of the United States of America}

\maketitle

\begin{article}

\begin{abstract}
The cold dark matter (CDM) cosmological model has been remarkably successful
in explaining cosmic structure over an enormous span of redshift,
but it has faced persistent challenges from observations that probe the 
innermost regions of dark matter halos and the properties of the Milky Way's 
dwarf galaxy satellites.  We review the current observational and theoretical 
status of these ``small scale controversies.'' Cosmological simulations that
incorporate only gravity and collisionless CDM predict halos with abundant 
substructure and central densities that are too high to match constraints
from galaxy dynamics.  The solution could lie in baryonic physics: recent 
numerical simulations and analytic models suggest that gravitational potential 
fluctuations tied to efficient supernova feedback can flatten the central cusps 
of halos in massive galaxies, and a combination of feedback and low 
star-formation efficiency could explain why most of the dark matter 
subhalos orbiting the Milky Way do not host visible galaxies. However, it is
not clear that this solution can work in the lowest mass galaxies where
discrepancies are observed. Alternatively, the small-scale conflicts could 
be evidence of more complex physics in the dark sector itself. For example, 
elastic scattering from strong dark matter 
self-interactions can alter predicted halo mass profiles, leading to good
agreement with observations across a wide range of galaxy mass. Gravitational
lensing and dynamical perturbations of tidal streams in the stellar halo 
provide evidence for an abundant population of low mass subhalos in accord with
CDM predictions. These observational approaches will get more powerful
over the next few years.
\end{abstract}






\section{Introduction}

The cold dark matter (CDM) hypothesis --- that dark matter consists
of a weakly interacting particle whose velocity dispersion in the
early universe was too small to erase structure on galactic or
sub-galactic scales --- emerged in the early 1980s and quickly became
a central element of the theory of cosmic structure formation.  Influential
early papers include Peebles' (1982) calculation of cosmic microwave
background (CMB) anisotropies and the matter power spectrum,
Blumenthal et al.'s (1984) investigation of galaxy formation in CDM,
and Davis et al.'s (1985) numerical simulations of galaxy clustering.
By the mid-1990s, the simplest CDM model with scale-invariant primordial
fluctuations and a critical matter density ($\Omega_m=1$) had
run afoul of multiple lines of observational evidence, including
the shape of the galaxy power spectrum, estimates of the mean
matter density from galaxy clusters and galaxy motions,
the age of the universe inferred from estimates of the Hubble constant,
and the amplitude of matter clustering extrapolated forward from the
fluctuations measured in the CMB.  Many variants on ``canonical''
CDM were proposed to address these challenges, and by the turn of
the century the combination of supernova evidence for cosmic acceleration
and CMB evidence for a flat universe
had selected a clear winner: \lcdm, incorporating cold dark
matter, a cosmological constant ($\Lambda$), and inflationary
initial conditions.  Today, the \lcdm\ scenario has a wide range
of observational successes, from the CMB to the Lyman-$\alpha$
forest to galaxy clustering to weak gravitational lensing,
and it is generally considered the ``standard model'' of cosmology.

However, as the resolution of cosmological N-body simulations improved in the
mid-to-late 1990s, they revealed two tensions with observations
that have remained thorns in the side of the CDM hypothesis.
First, simulations show that CDM collapse leads to cuspy
dark matter halos whose central density profiles rise as $r^{-\beta}$
with $\beta \sim 1 - 1.5$, while observed galaxy rotation curves
favor constant density cores in the dark matter distribution
(Flores \& Primack 1994; Moore 1994;
Navarro, Frenk, \& White 1997, hereafter NFW; Moore et al.\ 1999a).
Second, simulated halos retain a large amount
of substructure formed by earlier collapses on smaller scales,
predicting hundreds or thousands of subhalos in contrast to the $\sim 10$
``classical'' satellites of the Milky Way
(Klypin et al.\ 1999; Moore et al.\ 1999b).  These two conflicts
are often referred to as the ``cusp-core problem'' and the
``missing satellites problem.''   We will argue below
that these two problems have largely merged into one, and that
the most puzzling aspect of the Milky Way's satellite galaxies
is not their number but their low central matter densities,
which again imply mass profiles shallower than the naive CDM prediction.

In this brief article, based on our panel discussion at the
2012 Sackler Symposium on Dark Matter, we attempt to summarize the current
state of the CDM controversies at a level that will be useful
to those not immersed in the field.  The key question is whether
the conflicts between N-body predictions and observed galaxy
properties can be resolved by ``baryonic physics'' --- gas cooling,
star formation, and associated feedback --- or whether they
require different properties of the dark matter itself.

\section{Cores, Cusps, and Satellites}

\begin{figure*}[t]
\begin{center}
\resizebox{\columnwidth}{!}{\includegraphics{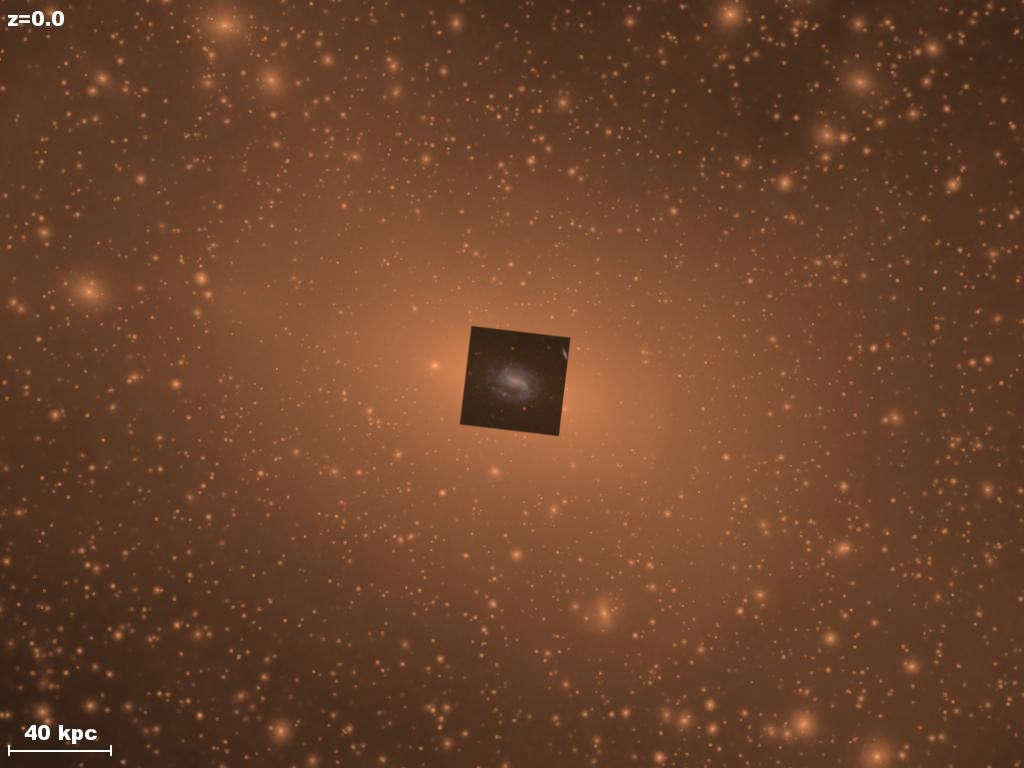}}
\hfill
\resizebox{\columnwidth}{!}{\includegraphics{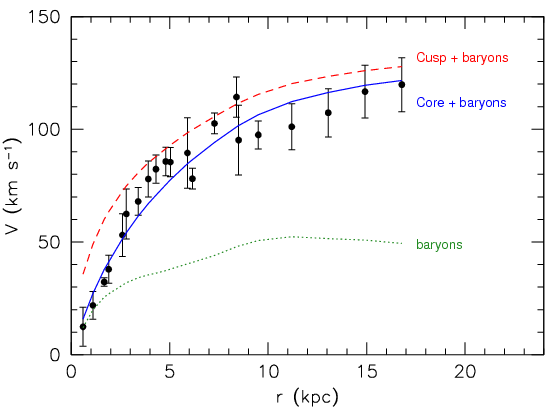}}
\end{center}
\caption{
The cusp-core problem.
{\it (Left)} An optical image of the galaxy F568-3 (small inset,
from the Sloan Digital Sky Survey) 
is superposed on the the dark matter distribution from
the ``Via Lactea'' cosmological simulation of a Milky Way-mass cold dark
matter halo (Diemand et al.\ 2007).  In the simulation image,
intensity encodes the square of
the dark matter density, which is proportional to annihilation rate
and highlights low mass substructure. 
{\it (Right)} The measured rotation curve of F568-3 (points)
compared to model fits assuming a
cored dark matter halo (blue solid curve) or a cuspy dark matter
halo with an NFW profile (red dashed curve, 
concentration $c=9.2$, $V_{200}=110\kms$).
The dotted green curve shows the contribution of baryons (stars+gas)
to the rotation curve, which is included in both model fits.
An NFW halo profile overpredicts the rotation speed in the inner
few kpc.  Note that the rotation curve is measured over roughly the scale
of the 40 kpc inset in the left panel.
}
\label{fig:kuzio}
\end{figure*}

Figure~1 illustrates the ``cusp-core'' problem.
To set the scene, the left panel superposes an optical image 
of the low surface brightness galaxy F568-3 onto a numerical simulation
of a cold dark matter halo.  The inner mass profile of a dark matter
halo can be probed by rotation curve measurements; 
for circular motions in a spherical
matter distribution, the rotation speed is simply
$v_c(r) = \sqrt{GM(r)/r}$, where $M(r)$ is the mass
interior to radius $r$.  
Points with error bars show the measured rotation curve of F568-3
(from Kuzio de Naray et al.\ 2008).
The dotted curve shows the $v_c(r)$
expected from the gravity of the stellar and gas components
of the galaxy, which are subdominant even in the central regions.
The solid curve shows the predicted rotation curve including the contribution
of an isothermal dark matter halo with a constant density core,
which fits the data well.  The dashed curve instead incorporates
a halo with an NFW profile and a concentration typical for
galaxy mass halos.  When normalized to match the observed rotation at
large radii, the NFW halo overpredicts the rotation speed in the
inner few kpc, by a factor of two or more.

Early theoretical discussions of the cusp-core problem devoted
considerable attention to the predicted central slope of
the density profiles and to the effects of finite numerical
resolution and cosmological parameter choices on the simulation
predictions (see Ludlow et al.\ 2013 for a recent,
state-of-the-art discussion).
However, the details of the profile shape are not essential
to the conflict; the basic problem is that CDM predicts too
much dark matter in the central few kpc of typical galaxies,
and the tension is evident at scales where $v_c(r)$ has
risen to $\sim 1/2$ of its asymptotic value
(see, e.g., Alam, Bullock, \& Weinberg 2002; Kuzio de Naray \& Spekkens 2011).
On the observational side, the most severe discrepancies
between predicted and observed rotation curves arise for fairly
small galaxies, and early discussions focused on whether
beam smearing or non-circular motions could artificially
suppress the measured $v_c(r)$ at small radii.
However, despite uncertainties in individual cases, improvements
in the observations, sample sizes, and modeling have led to
a clear overall picture: a majority of galaxy rotation curves
are better fit with cored dark matter profiles than with NFW-like
dark matter profiles, and some well observed galaxies cannot be fit
with NFW-like profiles, even when one allows halo concentrations at
the low end of the theoretically predicted distribution and accounts
for uncertainties in modeling the baryon component
(e.g., Kuzio de Naray et al.\ 2008).
Resolving the cusp-core problem therefore requires modifying
the halo profiles of typical spiral galaxies away from the profiles
that N-body simulations predict for collisionless CDM.

Figure~2 illustrates the ``missing satellite'' problem.
The left panel shows the projected dark matter density distribution
of a $10^{12} M_\odot$ CDM halo formed in a cosmological N-body
simulation.
Because CDM preserves primordial fluctuations down to very small
scales, halos today are filled with enormous numbers of subhalos
that collapse at early times and preserve their identities after
falling into larger systems.  Prior to 2000, there were only nine
dwarf satellite galaxies known within the $\sim 250$ kpc virial radius
of the Milky Way halo (illustrated in the right panel),
with the smallest having stellar velocity dispersions $\sim 10\kms$.
Klypin et al.\ (1999) and Moore et al.\ (1999b) predicted a factor
$\sim 5-20$ more subhalos above a corresponding velocity threshold
in their simulated Milky Way halos.
Establishing the ``correspondence'' between satellite stellar dynamics
and subhalo properties is a key technical point (Stoehr et al.\ 2002),
which we will return to below, but a {\it prima facie} comparison suggests
that the predicted satellite population far exceeds the observed one.

\begin{figure*}[t]
\begin{center}
\resizebox{5.0truein}{!}{\includegraphics{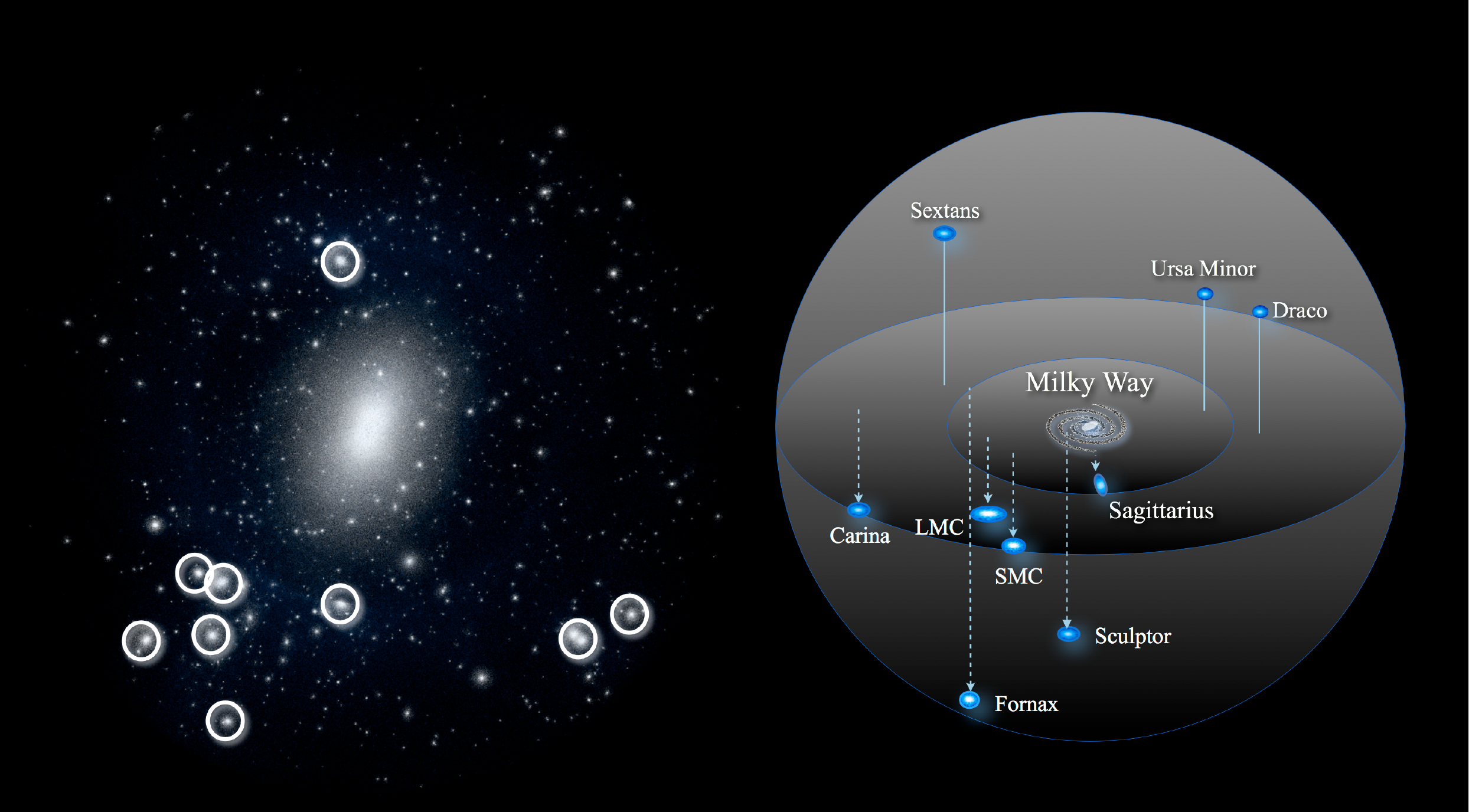}}
\end{center}
\caption{
The missing satellite and ``too big to fail'' problems.
{\it (Left)} Projected dark matter distribution (600 kpc on a side)
of a simulated, 
$10^{12} M_\odot$ CDM halo (Garrison-Kimmel, Boylan-Kolchin, \& Bullock, 
in preparation).  As in Figure~1, the numerous small subhalos far
exceed the number of known Milky Way satellites.
Circles mark the nine most massive subhalos.  
{\it (Right)} Spatial distribution of the ``classical'' satellites
of the Milky Way.  The central densities of the subhalos in the left panel
are too high to host the dwarf satellites in the right panel, predicting
stellar velocity dispersions higher than observed.
The diameter of the outer sphere in the right panel is 300 kpc; relative
to the simulation prediction (and to the Andromeda galaxy) the Milky Way's
satellite system is unusually centrally concentrated (Yniguez et al.\ 2013).
}
\label{fig:bullock}
\end{figure*}

Fortunately (or perhaps unfortunately), the missing satellite problem
seems like it could be solved fairly easily by baryonic physics.
In particular, the velocity threshold at which subhalo and dwarf satellite
counts diverge is close to the $\sim 30\kms$ value at which heating
of intergalactic gas by the ultraviolet photoionizing background
should suppress gas accretion onto halos,
which could plausibly cause these halos to remain dark
(Bullock, Kravtsov, \& Weinberg
2000; Benson et al.\ 2002; Somerville 2002).
Alternatively, supernovae and stellar winds from the first generation
of stars could drive remaining gas out of the shallow potential wells
of these low mass halos.  Complicating the situation, searches
using the Sloan Digital Sky Survey have discovered another $\sim 15$
``ultra-faint'' satellites
with luminosities of only $10^3 - 10^5 L_\odot$
(e.g., Willman et al.\ 2005; Belokurov et al.\ 2007).
The high-latitude SDSS imaging covered only $\sim 20\%$ of the sky, and many
of the newly discovered dwarfs are so faint that they could only be
seen to 50-100 kpc (Koposov et al.\ 2008; Walsh et al.\ 2009),
so extrapolating to the full volume within the Milky Way virial
radius suggests a population of several hundred faint dwarf satellites
(Tollerud et al.\ 2008).  Estimates from stellar dynamics
imply that the mass of dark matter in the central 0.3 kpc of
the host subhalos is $M_{0.3} \approx 10^7 M_\odot$ across an enormous
range of luminosities, $L \sim 10^3 - 10^7 L_\odot$ (encompassing the
``classical'' dwarf spheroidals as well as the SDSS dwarfs), which
suggests that the mapping between halo mass and luminosity becomes
highly stochastic near this mass threshold (Strigari et al.\ 2008).
The luminosity function of the faint and ultra-faint dwarfs can be
explained by semi-analytic models invoking photoionization and
stellar feedback (e.g., Koposov et al.\ 2009; Macci{\`o} et al.\ 2009),
though the efficiency of converting baryons to stars remains surprisingly
low ($\sim 0.1\%-1\%$) well above the photoionization threshold,
and it is unclear which if any of the ultra-faint dwarfs are
``fossils'' from before the epoch of reionization (Bovill \& Ricotti 2009).
Despite the gaps in understanding,
it seems reasonable for now to regard the relation between low
mass subhalos and ultra-faint dwarfs as a puzzle of galaxy formation
physics rather than a contradiction of CDM.

Instead, attention has focused recently on the most luminous satellites.
Circles in Figure~2 mark the nine most massive subhalos in the simulation,
which one would expect to host galaxies like the Milky Way's
``classical'' dwarf satellites.  However, the mass in the central regions
of these subhalos exceeds the mass inferred from
stellar dynamics of observed dwarfs, by a factor $\sim 5$
(Boylan-Kolchin et al.\ 2011, 2012; Springel et al. 2008;
Parry et al.\ 2012).
While it is possible in principle that these massive subhalos are dark
and that the observed dwarfs reside in less massive hosts, this
outcome seems physically unlikely; in the spirit of the times,
Boylan-Kolchin et al.\ (2011) titled this conflict ``too big to fail.''
The degree of discrepancy varies with the particular realization of
halo substructure and with the mass of the main halo,
but even for a halo mass at the low
end of estimates for the Milky Way the discrepancy appears too
large to be a statistical fluke, and a similar conflict is found
in the satellite system of the Andromeda galaxy
(Tollerud et al.\ 2012).  While ``missing satellites'' in low mass
subhalos may be explained by baryonic effects,
the ``too big to fail'' problem arises in more massive systems
whose gravitational potential is dominated by dark matter.  In its
present form, therefore, the satellite puzzle looks much like the
cusp-core problem: numerical simulations of CDM structure formation
predict too much mass in the central regions of halos and subhalos.
Indeed, Walker  \& Pe{\~n}arrubia (2011), Amorisco et al.\ (2013), and others
have reported evidence that the Milky Way satellites Fornax and Sculptor
have cored density profiles.


\section{Solutions in Baryonic Physics?}

When the cusp-core problem was first identified, the
conventional lore was that including baryonic physics would only
exacerbate the problem by adiabatically contracting the dark matter
density distribution (Blumenthal et al.\ 1986; Flores \& Primack
1994).  Navarro, Eke, \& Frenk (1996) proposed a scenario, which
seemed extreme at the time, for producing a cored dark matter
distribution: dissipative baryons draw in the dark matter orbits
adiabatically by slowly deepening the gravitational potential, then
release them suddenly when the supernova feedback of a vigorous
starburst blows out a substantial fraction of the baryonic material,
leaving the dark matter halo less concentrated than the one that would
have formed in the absence of baryons.  Since then, hydrodynamic
simulations have greatly improved in numerical resolution and in the
sophistication with which they model star formation and supernova
feedback.  With the combination of a high gas density threshold for 
star formation and efficient feedback, simulations successfully
reproduce the observed stellar and cold gas fractions of field galaxies.
The ejection of low angular momentum gas by feedback plays a critical
role in suppressing the formation of stellar bulges in dwarf galaxies
(Governato et al.\ 2010), another long-standing problem in early
simulations of galaxy formation.  The episodic gas outflows also  produce rapid
fluctuations of the gravitational potential, in contrast to the steady
growth assumed in adiabatic contraction models.

\begin{figure*}[t]
\begin{center}
\hfil
\resizebox{!}{3.0truein}{\includegraphics{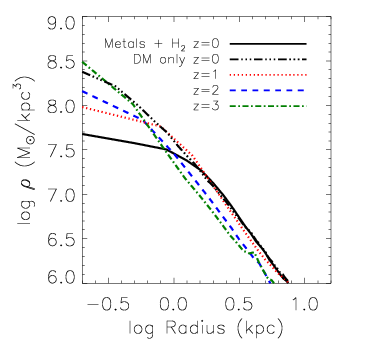}}
\hskip 0.2truein
\resizebox{!}{3.0truein}{\includegraphics{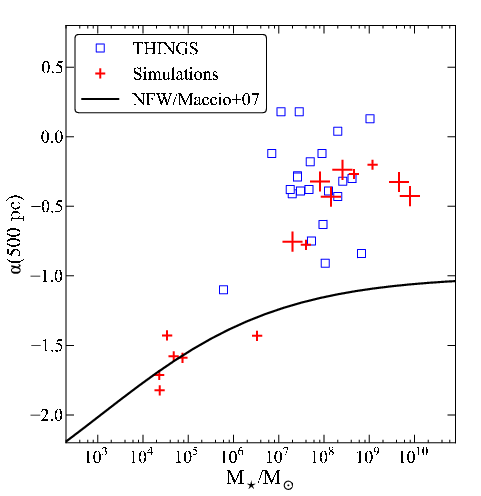}}
\hfil
\end{center}
\caption{
Baryonic effects on CDM halo profiles in cosmological simulations,
from Governato et al.\ (2012).
{\it (Left)} The upper, dot-dash curve shows the cuspy dark matter density
profile resulting from from a collisionless N-body simulation.
Other curves show the evolution of the dark matter profile in a 
simulation from the same initial conditions that includes gas
dynamics, star formation, and efficient feedback.  By $z=0$ (solid
curve) the perturbations from the fluctuating baryonic potential
have flattened the inner profile to a nearly constant density core.
{\it (Right)} Logarithmic slope of the dark matter profile $\alpha$
measured at 0.5 kpc, as a function of galaxy stellar mass.
Crosses show results from multiple hydrodynamic simulations.
Squares show measurements from rotation curves of observed galaxies.
The black curve shows the expectation for pure dark matter simulations,
computed from NFW profiles with the appropriate concentration.
For $M_* > 10^7 M_\odot$, baryonic effects reduce the halo
profile slopes to agree with observations.
}
\label{fig:fabio}
\end{figure*}

Figure~3, based on Governato et al.\ (2012), illustrates the
impact of this episodic feedback on the dark matter density profile.
In the left panel,
the upper dot-dashed curve shows the final halo profile of an N-body
simulation run with gravity and dark matter only.  Other curves show
the evolution of the {\it dark matter} density profile in a
hydrodynamic simulation with star formation and feedback, from the
same initial conditions.  Over time, the central dark matter density
drops, and the cuspy profile is transformed to one with a nearly
constant density core (lower solid curve).  
Pontzen \& Governato (2012) present an
analytic model that accurately describes this transformation (and its
dependence on simulation assumptions); essentially, the rapid
fluctuations in the central potential pump energy into the dark matter
particle orbits, so that they no longer penetrate to the center of the
halo.  The Governato et al.\ simulations use smoothed particle
hydrodynamics, and the same flattening of dark matter cusps is found
in adaptive mesh refinement simulations that have similarly episodic
supernova feedback (Teyssier et al. 2012).  

The right panel of Figure~3 compares the density profile slopes of
simulated galaxies to observational estimates from 21cm measurements
of nearby galaxies (Walter et al.\ 2008) and to predictions for
an NFW dark matter halo.  The reduced central density slopes
agree well with observations for galaxies with stellar mass
$M_* > 10^7 M_\odot$.  Strong gas outflows are observed in a wide
variety of galaxies, including the likely progenitors of 
$M_* \sim 10^8-10^9 M_\odot$ dwarfs observed at $z \sim 2$
(van der Wel et al.\ 2011).  However, for galaxies with 
$M_*$ below $\sim 10^7 M_\odot$, analytic models suggest that with so few
stars there is not enough energy in supernovae alone to create 
dark matter cores of $\sim 1$ kpc (Pe\~narrubia et al.\ 2012).
More generally, Garrison-Kimmel et al.\ (2013) used idealized,
high resolution 
simulations to model potential fluctuations of the type
expected in episodic feedback models and concluded that the 
energy required for solving the ``too big to fail'' problem
exceeds that available from supernovae in galaxies with stellar
masses below $\sim 10^7 M_\odot$.
The low mass galaxies in Figure~3 (from Governato et al.\ 2012)
are consistent with this expectation, with
density profile slopes that are negligibly affected by feedback
at the 0.5 kpc scale.  On the other hand, high resolution simulations
of luminous {\it satellites} in the halo of Milky Way-like hosts do show
reduced central dark matter densities, from a combination of early
feedback effects with ram pressure stripping and tidal heating
by the host halo and disk, processes that can
extract energy from the host galaxy's gravitational potential
(Arraki et al.\ 2012; Zolotov et al.\ 2012; Brooks et al.\ 2013).
Alternatively, Kuhlen et al.\ (2013) argue that molecular cooling physics
may make star formation efficiency highly stochastic at a halo mass as high
as $10^{10} M_\odot$, so that even the Milky Way's most massive subhalos
are not ``too big to fail.''  Ram pressure in the Galactic halo could then
remove the gas from the dark subhalos.

These arguments point to {\it isolated},
low-mass galaxies with $M_* \sim 10^6-10^7 M_\odot$ as ideal
laboratories for testing the predictions of CDM-based models.
Dwarfs that are far separated from a giant galaxy must rely on their own
(modest) supernova reservoirs for energy injection.
Ferrero et al. (2012) have studied a population of 
$\sim 10^6 - 10^7 M_\odot$ field galaxies and argued that the central density 
problem persists even for relatively isolated dwarfs of this size.  If
this result holds up in further investigations, it will become a particularly
serious challenge to CDM.

\section{Solutions in Dark Matter Physics?}


Instead  of complex baryonic effects, the cusp-core and satellite
problems could indicate a failure of the CDM hypothesis itself.
One potential solution is to make dark matter ``warm,'' so that
its free-streaming velocities in the early universe are large
enough to erase primordial fluctuations on sub-galactic scales.
For a simple thermal relic, the ballpark particle mass is $m \approx 1\,$keV,
though details of the particle physics can alter the relation
between mass and the free-streaming scale, which is the important
quantity for determining the fluctuation spectrum.
Alternatively, the small scale fluctuations can be suppressed
by an unusual feature in the inflationary potential
(Kamionkowski \& Liddle 2000).  While collisionless collapse
of warm dark matter (WDM) still leads to a cuspy halo profile,
the central concentration is lower than that of CDM halos when the
mass scale is close to the spectral cutoff (e.g., Avila-Reese
et al.\ 2001), thus allowing a better fit to observations of galaxy rotation
curves and dwarf satellite dynamics.  The mass function of
halos and subhalos drops at low masses because there are
no small scale perturbations to produce collapsed objects,
so the subhalo mass function can be brought into agreement
with dwarf satellite counts.  There have been numerous
numerical simulations of structure formation with WDM;
recent examples include Polisensky \& Ricotti (2011), 
Anderhalden et al. (2012), Lovell et al. (2012), 
Macci{\`o} et al. (2012), Schneider et al. (2012), and Angulo et al. (2013).

Warm dark matter is a ``just-so'' solution to CDM's problems,
requiring a particle mass (or free-streaming velocity) that is tuned
to the particular scale of dwarf galaxy halos.
However, the more serious challenge to WDM is
observational, for two reasons.  First, WDM does too good a job in
eliminating power on small scales; for a thermal relic of mass $m = 2\hbox{
keV}$, there are too few subhalos in the Milky Way to host the known satellite
galaxies (Polisensky \& Ricotti 2011).  It also appears in conflict with
observations of strong-lens systems, which show evidence for a significant
subhalo fraction as well as the existence of small ($10^8M_\odot$) subhalos
(Dalal \& Kochanek 2002; Dobler \& Keeton 2006; Vegetti et al. 2010a,b, 2012;
Fadely \& Keeton 2011, 2012).  Second, suppressing primordial fluctuations on
small scales alters the predicted structure of
Lyman-$\alpha$ forest absorption towards quasars at high redshift,
where these scales are still in the quasi-linear regime
(Narayanan et al.\ 2000).
Recent studies of the Lyman-$\alpha$ forest set a lower limit
on the dark matter particle mass of several keV, high enough
that the dark matter is effectively ``cold'' from the point of
view of the cusp-core problem
(Seljak et al.\ 2006; Viel et al.\ 2008; but see Abazajian 2006 for a
counterclaim of a lower minimum particle mass).
Even setting these problems aside, it appears that
WDM on its own does not fix the shape of rotation curves across
the full range of galaxy masses where conflict with CDM is observed
(Kuzio de Naray et al.\ 2010).
While some uncertainties in the numerical simulations and observational
data remain, it appears that WDM cannot solve the cusp-core 
and missing satellite problems while remaining consistent with 
Lyman-$\alpha$ forest and substructure observations.

\begin{figure*}[t]
\begin{center}
\hfil
\resizebox{2.0truein}{!}{\includegraphics{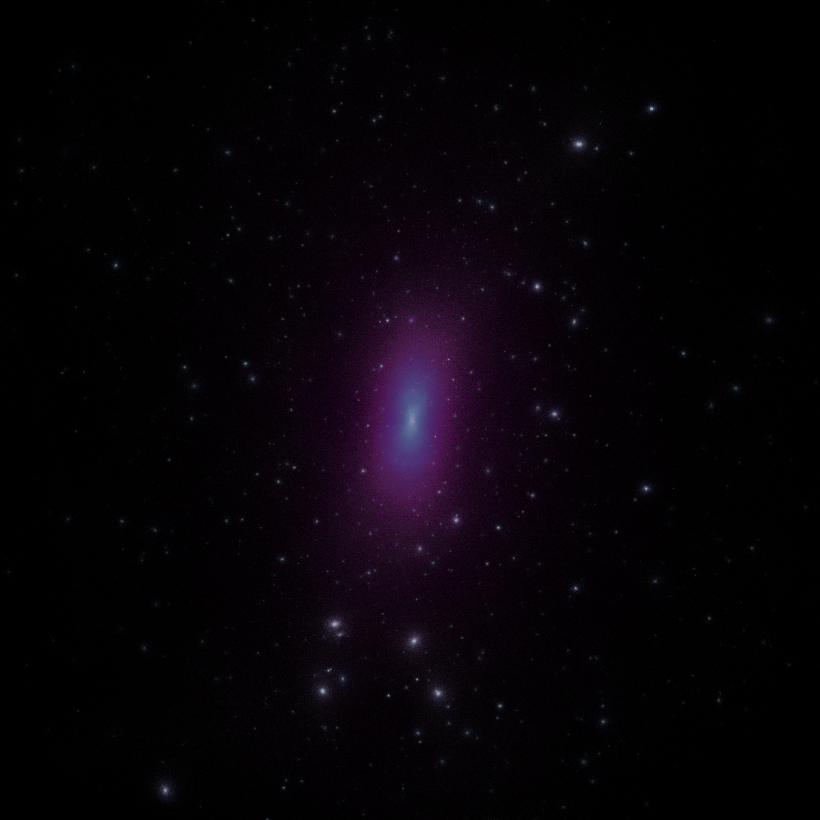}}
\resizebox{2.0truein}{!}{\includegraphics{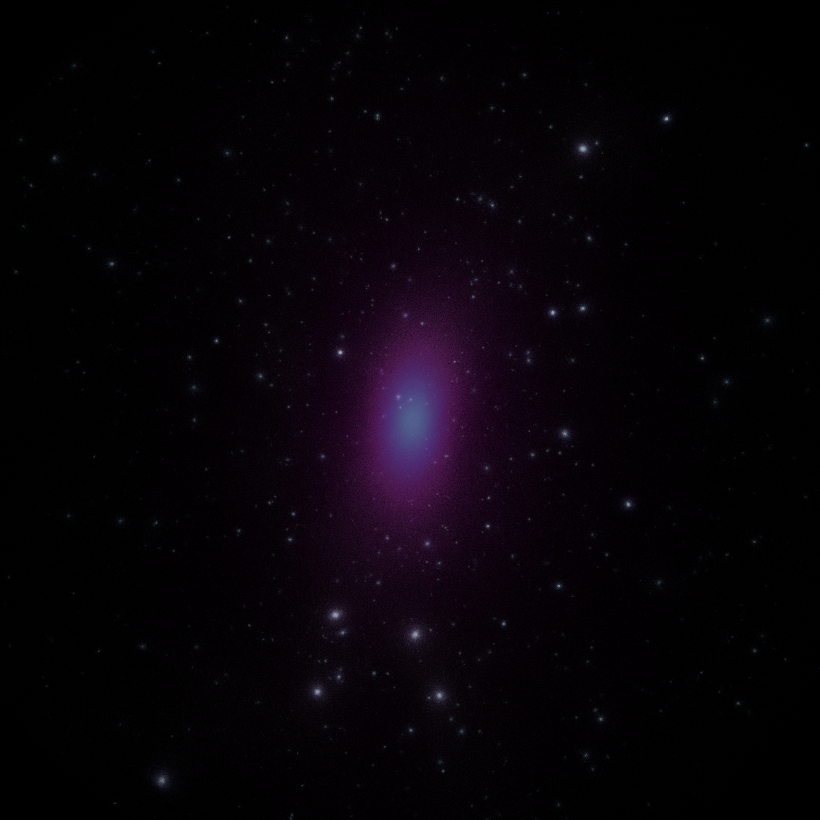}}
\resizebox{2.5truein}{!}{\includegraphics{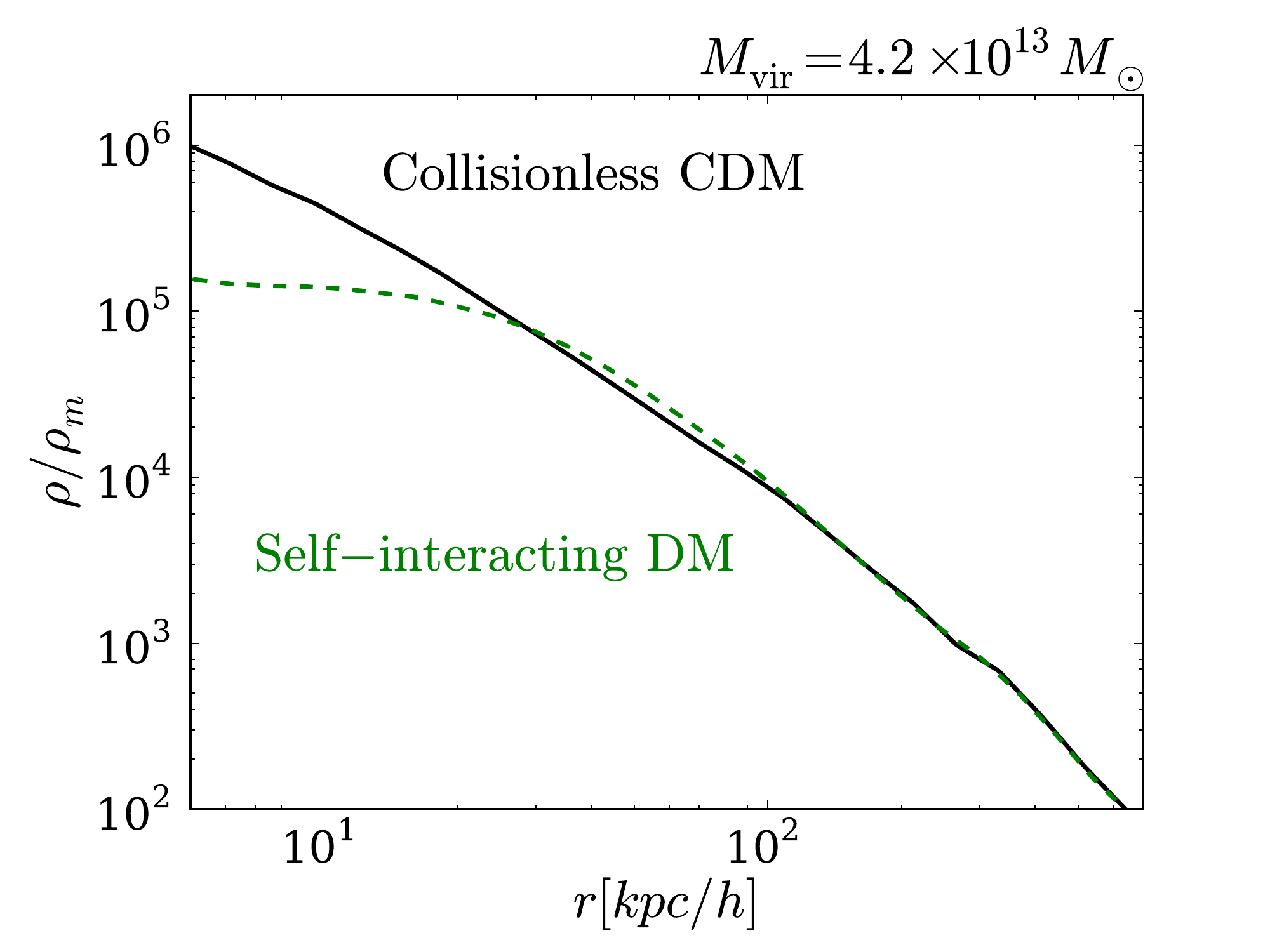}}
\hfil
\end{center}
\caption{
Effect of self-interacting dark matter (SIDM) on halo structure, from
simulations by Rocha et al.\ (2013).  The left panel shows a Milky Way mass
CDM halo, and the middle panel shows the same halo from an SIDM simulation
with cross-section of $1\cgm$.  The structure and substructure are similar,
but the SIDM halo is rounder and less dense in the center.
The right panel compares the density profiles of a CDM and SIDM halo, 
showing the core produced by elastic scattering.  This halo has 
$M = 4.2\times 10^{13} M_\odot$, but similar behavior is found at other
halo masses.
}
\label{fig:peter}
\end{figure*}

An alternative idea, made popular by Spergel \& Steinhardt (2000),
is that cold dark matter has weak interactions with baryons
but strong self-interactions.  The required scattering cross-section
is roughly $(m/{\rm g})^{-1}\,{\rm cm}^{2}$ where $m$
is the particle mass; note that $1\cgm \approx 1\,{\rm barn}\,{\rm GeV}^{-1}$ 
is approximately a nuclear-scale cross section.
In this case, elastic scattering in the dense central regions
of halos is frequent enough to redistribute energy and angular momentum
among particles, creating
an isothermal, round core of approximately constant density (Burkert 2000).
Some early studies suggested that this idea was ruled out
by gravitational lensing (Miralda-Escud\'e 2002) or by catastrophic
gravitational core collapse found in a simulation of an isolated halo
(Kochanek \& White 2000), but recent numerical studies show
that these concerns are not borne out in fully cosmological simulations.
Instead, simulations show that there is a viable window of mass
and cross-section
where self-interacting dark matter (SIDM) can produce
cored dark matter profiles and remain consistent with
observational constraints (Rocha et al. 2013; Peter et al. 2013).  

Figure~4, based on Rocha et al.\ (2013),
compares the structure and density profiles of halos formed from the same
initial conditions with collisionless CDM and SIDM.
Elastic scattering in the central regions, where an average particle
experiences a few collisions per Hubble time, flattens the density cusp
and reduces triaxiality.
The scattering mechanism would operate across a wide range
of halo masses, allowing SIDM to address both the rotation curves
of Milky Way-like galaxies and the central densities of
dwarf satellites.  Because they are more weakly bound,
SIDM subhalos are more easily subject to tidal disruption
than CDM subhalos. However, the suppression of the low-mass subhalo
count is not significant for allowed cross sections except in the innermost
region of the host halo (Vogelsberger et al. 2012; Rocha et al 2013).  Thus,
SIDM can solve the cusp-core problem while leaving enough subhalos to host
Milky Way satellites, unlike WDM.

The prospects for SIDM appear much more hopeful than for WDM
(though for a summary of pro-WDM views see Biermann et al.\ 2013).
Velocity-independent cross sections in the range 
$\sim 0.1-0.5\cgm$
create cores that are approximately the right size for Milky Way
dwarf galaxies, spiral galaxies, and galaxy clusters (Newman et al. 2013a,b;
Rocha et al. 2013) while leaving halos triaxial enough to match observations
(Peter et al. 2013).  Cross sections in this range are also consistent with
observations of merging galaxy clusters (Clowe et al. 2006;
Randall et al. 2008; Dawson et al. 2012).  Moreover, particle model builders
have recently focused attention on new classes of ``hidden sector'' models
that generically produce SIDM particle candidates, although in general the
elastic scattering cross section has a strong velocity dependence (Ackerman et
al. 2009; Buckley 2010; Feng et al. 2010; Tulin et al. 2013a,b).  For these
models, strong self-interactions may only be present in a narrow range of
halo mass, leaving halos on other scales effectively collisionless.
Observationally, the goal is to either rule out or find evidence for
SIDM cross sections $\sigma > 0.1\cgm$, as for smaller cross-sections the
halo phenomenology is likely to be indistinguishable from CDM.

There are alternative dark matter physics mechanisms that
could reduce the central densities of halos, including
particle decay and particle-antiparticle
annihilation (Kaplinghat et al. 2000; Peter et al. 2010)
or the recently suggested possibility
of escape from flavor-mixed quantum states (Medvedev 2013).

\section{Conclusions}

Are the tensions between CDM predictions and observations
on the scales of galactic cores and satellite halos telling
us something about the fundamental properties of dark matter,
or are they telling us something interesting about the complexities
of galaxy formation?  After two decades of debate,
the current state of the field is an
unsatisfying stalemate (or, perhaps, a draw by repetition).
However, there are several directions for future progress that
could resolve the question.

Developments of the last several years have focused the
``small scale controversies'' down to one fairly specific issue, the
influence of baryons on the dark matter halo profile in systems
where the baryons are today greatly sub-dominant.  A variety of
studies have shown that baryonic effects can plausibly account
for cores in halos occupied by high surface-brightness galaxies
and can plausibly suppress star formation in very low mass halos.
Improved simulations may show that baryonic effects can soften
cusps even in galaxies that are now dark matter dominated, or
they may show that the energetics arguments summarized above
do indeed point to a genuine problem for CDM that cannot be
resolved by supernova feedback or galactic tides.
Improved simulations of models with interacting dark matter
may show that they can readily solve the small scale problems,
or they may show that cross-section parameters chosen to match
one set of observations ultimately fail when confronted with another set.
SIDM models might also be ruled out if they predict
halo shapes that can be excluded by observations of
stellar or gas dynamics.  Improved measurements of stellar velocities
in satellite galaxies, and discovery
of new satellites from imaging surveys such as Pan-STARRS
and the Dark Energy Survey, may better delineate the satellite problem
itself.

These developments will affect the credibility of baryonic and dark matter
solutions to the CDM controversies, but they may not yield a
definitive conclusion.  More satisfactory would be a direct test of
the CDM prediction that vast numbers of low mass subhalos
($\sim 20,000$ with masses $>10^6 M_\odot$, and $\sim 2,000,000$ above
$10^4 M_\odot$) are orbiting within the virial radius of the Milky Way
and similar galaxies.  Flux anomalies in gravitational lenses
have already provided important evidence for subhalos that collectively 
contain a few
percent of the mass within their parent halos, a level roughly consistent
with CDM predictions and an order of magnitude above that expected from
luminous satellites alone (Dalal \& Kochanek 2002; Metcalf \& Amara 2012).
These anomalies do not directly probe the mass spectrum of the subhalos,
though at masses $M \sim 10^8 M_\odot$ they produce detectable astrometric 
deviations in addition to flux anomalies (e.g., Vegetti et al.\ 2012).
Current constraints are derived from a small sample of lensed radio-loud
quasars, as optical flux anomalies could be produced by stellar microlensing.
Statistics should improve dramatically after the launch of the 
{\it James Webb Space Telescope}, which can resolve lenses at mid-IR
wavelengths that are not affected by stellar microlensing because the quasar
dust emission regions are too large.
An alternative route is to study cold tidal streams in the
Milky Way, which would be perturbed by the multitude of
passing subhalos.  Carlberg \& Grillmair (2013, and references
therein) argue that observed tidal streams already show evidence of
these perturbations, and a combination of better numerical simulations,
more streams, and more detailed density and dynamical measurements could
yield definitive evidence for or against CDM's predicted
subhalo population.  Further theoretical work is needed to determine whether
lensing or stream perturbations can distinguish CDM from SIDM.

As emphasized throughout the Sackler Symposium, there are great hopes
that underground detection experiments, $\gamma$-ray observations,
or collider experiments will identify the dark matter particle
within the next decade.  Such detections might definitively
demonstrate whether dark matter is cold and weakly interacting,
or they might unmask the particle while yielding
ambiguous answers to this question.
In the meantime, astronomers will continue their decades-long
practice of studying the dark sector by observing the visible.





\begin{acknowledgments}
We gratefully acknowledge our many collaborators whose work we have 
summarized here.
We thank Chris Kochanek for helpful comments on lensing constraints.
Our work on these topics is supported by NASA and the NSF, including
grants NNX10AJ95G, NNX 08AG84G, AST-1009505, AST-1009973, and
AST-0607819.
\end{acknowledgments}





\end{article}








\end{document}